\documentclass[prl,twocolumn,showpacs,
superscriptaddress,floatfix]{revtex4}
\usepackage{graphicx}
\usepackage{latexsym}

\begin{document}

\title{Dynamic Correlation in Wave Propagation in Random Media}

\author{A.A.~Chabanov}
\affiliation{Department of Physics, Queens College of the City
University of New York, Flushing, NY 11367, USA}
\affiliation{CEMS, University of Minnesota, Minneapolis, MN 55455,
USA}

\author{B.~Hu}
\affiliation{Department of Physics, Queens College of the City
University of New York, Flushing, NY 11367, USA}

\author{A.Z.~Genack}
\affiliation{Department of Physics, Queens College of the City
University of New York, Flushing, NY 11367, USA}

\date{February 20, 2004}

\begin{abstract}
We report time-resolved measurements of the statistics of pulsed
transmission through quasi-one-dimensional dielectric media with
static disorder. The normalized intensity correlation function
with displacement and polarization rotation for an incident pulse
of linewidth $\sigma$ at delay time $t$ is a function only of the
field correlation function, which is identical to that found for
steady-state excitation, and of $\kappa_{\sigma}(t)$, the residual
degree of intensity correlation at points at which the field
correlation function vanishes. The dynamic probability
distribution of normalized intensity depends only upon
$\kappa_{\sigma}(t)$. Steady-state statistics are recovered in the
limit $\sigma$$\to$0, in which $\kappa_{\sigma=0}$ is the
steady-state degree of correlation.
\end{abstract}

\maketitle The statistics of steady-state classical wave
propagation \cite{Goodman,ShengBook,Rossum99} and electronic
conductance \cite{Mesobook} in disordered systems reflect the
superposition of partial waves following trajectories with a wide
distribution of path lengths, which is proportional to the
particle time-of-flight distribution. Since the lengths of
trajectories increase with time, paths cross upon themselves more
frequently and the impact of weak localization could be expected
to build in time
\cite{Altshuler88,Muzyk95,Mirlin00,Prigodin94,Weaver,Titov00,Schomerus01,Dynamics03,Haney03,Bart03,Dynamics04}.
This was demonstrated recently in the observation of an increase
of the coherent backscattering enhancement from 2 to 3 in
time-resolved acoustic measurements in a three-dimensional elastic
body \cite{Weaver}, and in a time-decaying leakage rate of
microwave radiation from a quasi-1D random dielectric sample
\cite{Dynamics03}. To achieve a systematic understanding of weak
localization in the time domain, it is essential to examine the
statistics of propagation in addition to ensemble-averaged
transport. This can be accomplished by parsing transmission
according to the delay from an exciting pulse and studying the
correlation and probability distribution of intensity as a
function of delay time, and finally by relating these dynamics to
steady-state statistics obtained under monochromatic excitation. A
dynamical perspective on the impact of gain or loss on
steady-state statistics might be of particular interest since the
distribution of trajectories within the sample at a given delay is
not altered by the presence of inelastic processes, while the
average temporal profile changes.

Nonlocal intensity correlation
\cite{thcorr,excorr,Patrick02,Polar04} leads to giant transmission
fluctuations
\cite{Rossum99,thcorr,excorr,Rossum,Kogan,Marin,Patrick02,Polar04,Yamilov04}
and lies at the heart of mesoscopic physics \cite{UCF}. The
intensity correlation function associated with bulk scattering
versus displacement or polarization rotation of a detector may be
expressed as $C$=$F$+$\kappa(F+1)$
\cite{Patrick02,Polar04,Yamilov04}, where $F$$=$$|F_E|^2$ is the
square of the field correlation function. The degree of
correlation, $\kappa$, is the value of $C$ when $F$ vanishes,
which occurs, for example, for a polarization shift of $90^0$ or
for displacements much greater than a wavelength. In the absence
of inelastic processes, the probability distributions of both
intensity and total transmission normalized to their respective
ensemble averages, $P(s_{ab})$ and $P(s_a)$, where $a$ and $b$ are
modes of the incident and transmitted waves, are obtained from
diagrammatic \cite{Rossum} and random matrix \cite{Kogan} theories
in terms of the dimensionless conductance, $g$ \cite{g}, in the
diffusive limit, $g$$\gg$1. For both quantum and classical waves,
$g$ is equal to the sum of transmission coefficients over all
input and output modes \cite{Landauer}. Surprisingly, theoretical
expressions for $P(s_{ab})$ and $P(s_a)$ \cite{Rossum,Kogan}
closely match the measured distributions \cite{Marin} even in the
presence of absorption, and even at the localization threshold,
reached when $g$$\sim$1, when $g$ is replaced by $2/3var(s_a)$.

In this Letter, we report microwave measurements of the
time-resolved field transmitted through random quasi-1D dielectric
samples. Remarkably, the field correlation functions with
displacement and polarization rotation at any time are identical
to those found in steady state. We also find that the
corresponding cumulant correlation functions of the normalized
intensity have the same dependence upon the field correlation
functions as in the frequency domain,
$C_{\sigma}(t)$=$F$+$\kappa_{\sigma}(t)(F+1)$, with a parameter,
$\kappa_{\sigma}(t)$, expressing the degree of correlation at time
$t$ following excitation by a Gaussian pulse of width $\sigma$. We
find further that the probability distribution of normalized
intensity, $P(s_{ab}(t))$, has the same form as for the
steady-state transmitted intensity distribution
\cite{Rossum,Kogan}, but with $2/3\kappa_{\sigma}(t)$ substituted
for $g$. We find that even in diffusive samples,
$\kappa_{\sigma}(t)$ reaches values exceeding the steady-state
value at the Anderson localization threshold \cite{Anderson} of
$\kappa$$\simeq$2/3 \cite{Nature}. Steady-state statistics are
found to be a limiting case of dynamic statistics, in which the
incident pulse linewidth vanishes, $\sigma$$\to$0, with
$\kappa_{\sigma=0}$=$\kappa$. These results show that
$\kappa_{\sigma}(t)$ is the essential function describing the
statistics of wave propagation.

Spectral measurements of the field transmission coefficient of
microwave radiation as a function of rotation of linear
polarization and displacement are made with use of a vector
network analyzer in ensembles of random dielectric samples in
which the wave is diffusive. The samples are contained in a copper
tube with open ends of length $L$ greatly exceeding its 7.3-cm
diameter. New sample realizations are produced by briefly rotating
the sample tube about its axis after each spectrum is taken. The
response to a pulse with a Gaussian temporal envelope at carrier
frequency $\nu_0$ is obtained by Fourier transforming the product
of the field transmission spectrum and a Gaussian spectral
function of width $\sigma$.

Polarization-selective measurements of the transmitted microwave
field are made with use of a conical horn detector \cite{Polar04}.
The samples are composed of 0.95-cm-diameter alumina spheres with
refractive index 3.14 embedded in Styrofoam spheres of diameter
1.9 cm and refractive index 1.04 at an alumina volume fraction of
0.068. Measurements are made in an ensemble of 10,000 alumina
configurations of $L$=61 cm (Sample A) over the frequency range
14.7-15.7 GHz in steps of 0.6 MHz for a single orientation of the
horn detector. Measurements are also carried out in an ensemble of
12,000 alumina configurations of $L$=90 cm (Sample B) over the
frequency range 16.95-17.05 GHz in 1-MHz steps, for 7 orientations
of the horn detector rotated in steps of $15^0$ over a $90^0$
range. Steady-state measurements of intensity correlation give
$\kappa$=0.09 in Sample A \cite{Dynamics03} and $\kappa$=0.29 in
Sample B \cite{Polar04}.

Examples of the response to excitation pulses with three different
values of $\sigma$ in a single random realization of Sample A are
shown in Fig.~1a versus time delay from the center of the incident
pulses in units of the diffusion time,
$t_D$=$(L$+$z_0)^2\!/\pi^2D$, where $D$ is the diffusion
coefficient and $z_0$ is the boundary extrapolation length. The
width of temporal intensity fluctuations is seen to be
approximately equal to that of the incident pulse. This can be
expressed quantitatively via the correlation function with time
shift $\Delta t$ of the transmitted field normalized to the square
root of the ensemble average of the time-varying intensity at time
$t$, $F_{E_{\sigma}}$=$\langle
E_{\sigma}^*(t)E_{\sigma}(t$+$\Delta t)\rangle/\!\sqrt{\langle
I_{\sigma}(t)\rangle\langle I_{\sigma}(t\!+\!\Delta t)\rangle}$.
We find that $F_{E_{\sigma}}$ is independent of $t$. In the case
of a Gaussian incident pulse of bandwidth $\sigma$,
$|F_{E_{\sigma}}|$ is a Gaussian with width
$(\sqrt{2}\pi\sigma)^{-1}$ (Fig.~1b). We also find that, though
$I_{\sigma}(t)$ in any given realization depends strongly upon
$\sigma$, $\langle I_{\sigma}(t)\rangle$ for $t>t_D$ depends only
weakly upon $\sigma$, once $\sigma>1/\pi^2t_D$.

\begin{figure}[t!]
\includegraphics[width=\columnwidth]{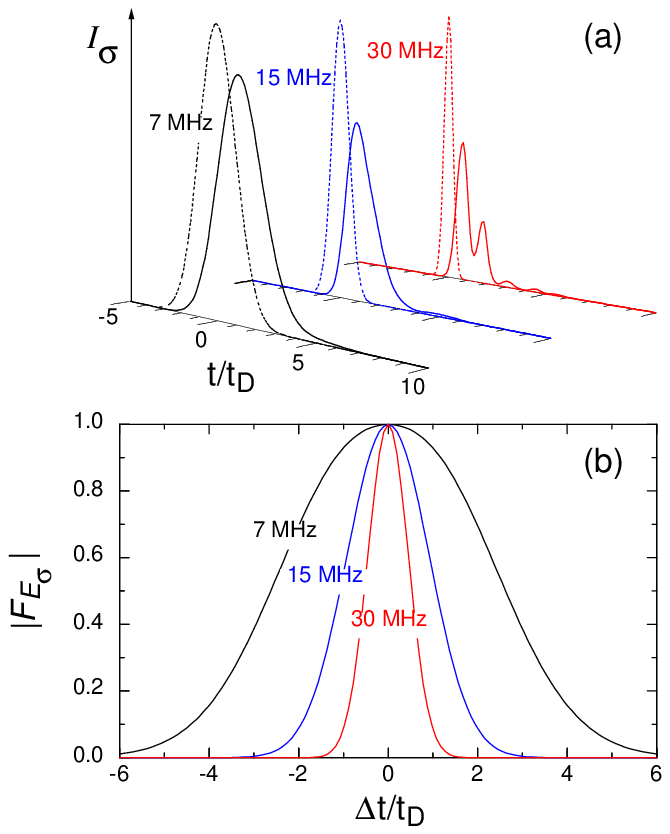}
\caption{(a) Transmitted intensity through a random realization of
Sample A (solid curves) following incident Gaussian pulses (dashed
curves) with $\sigma$=7, 15, and 30 MHz. The incident pulses are
centered at $t$=0 and shown with the same height. (b) Absolute
value of the field correlation function of
$E_{\sigma}(t)/\!\sqrt{\langle I_{\sigma}(t)\rangle}$ with time
shift for the values of $\sigma$ in (a).}
\end{figure}

\begin{figure}[t!]
\includegraphics[width=\columnwidth]{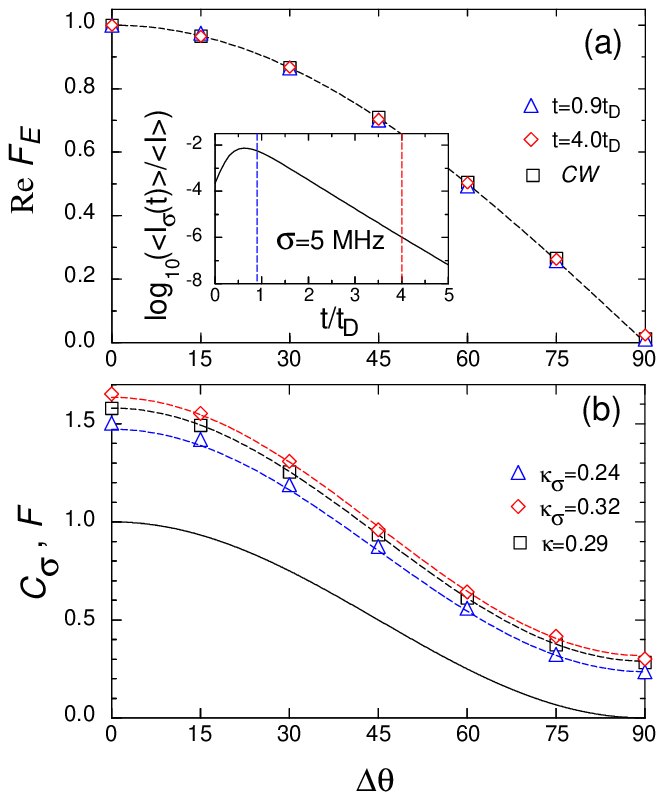}
\caption{(a) Real part of the field correlation function,
$F_E(\Delta\theta)$, and (b) intensity correlation function,
$C_{\sigma}(\Delta\theta)$, with polarization rotation of the
transmitted wave through Sample B at the two delay times following
pulsed excitation with $\sigma$=5 MHz and for monochromatic
excitation (\textit{CW}). The dashed curves are (a)
$F_E(\Delta\theta)$=$\cos(\Delta\theta)$ and (b)
$C_{\sigma}(\Delta\theta,t)$=$F(\Delta\theta)$+$\kappa_{\sigma}(t)[F(\Delta\theta)$+$1]$
with the values of $\kappa_{\sigma}(t)$ indicated. The solid curve
is $F(\Delta\theta)$. The logarithm of the average pulsed
transmission through Sample B for $\sigma$=5 MHz, normalized by
the average steady-state transmitted intensity, is shown in the
inset. The delay times at which correlation is presented in the
figure are indicated by vertical dashed lines.}
\end{figure}

\begin{figure}[t!]
\includegraphics[width=\columnwidth]{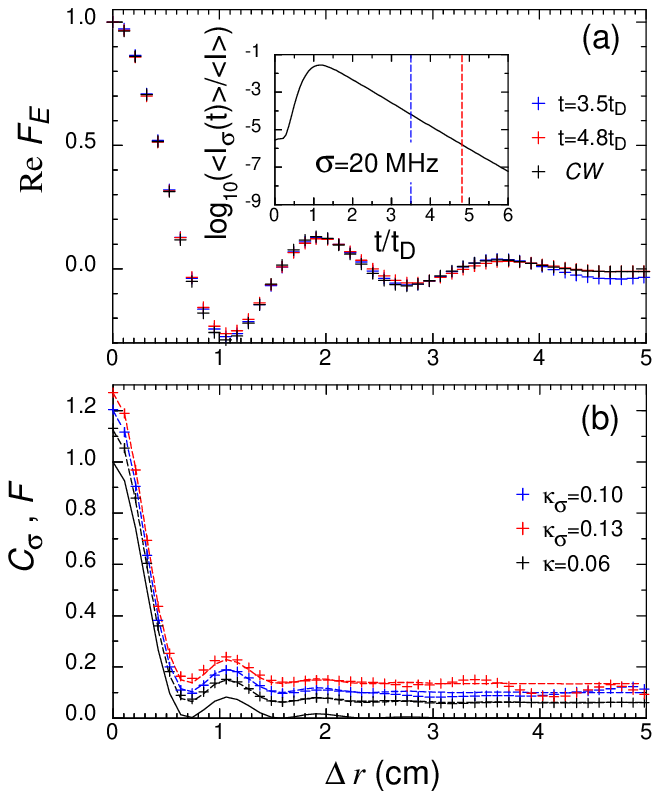}
\caption{(a) Real part of the field correlation function
$F_E(\Delta r)$, and (b) intensity correlation function,
$C_{\sigma}(\Delta r)$, with displacement on the output surface of
a random Polystyrene sample at the two delay times following
pulsed excitation with $\sigma$=20 MHz and for monochromatic
excitation (\textit{CW}). The dashed curves in (b) are
$C_{\sigma}(\Delta r,t)$=$F(\Delta
r)$+$\kappa_{\sigma}(t)[F(\Delta r)$+$1]$, with the values of
$\kappa_{\sigma}(t)$ indicated. The solid curve is $F(\Delta r)$.
The logarithm of the average pulsed transmission through the
sample for $\sigma$=20 MHz, normalized by the average steady-state
transmitted intensity, is shown in the inset. The delay times at
which correlation is presented in the figure are indicated by
vertical dashed lines.}
\end{figure}

\begin{figure}[t!]
\includegraphics[width=\columnwidth]{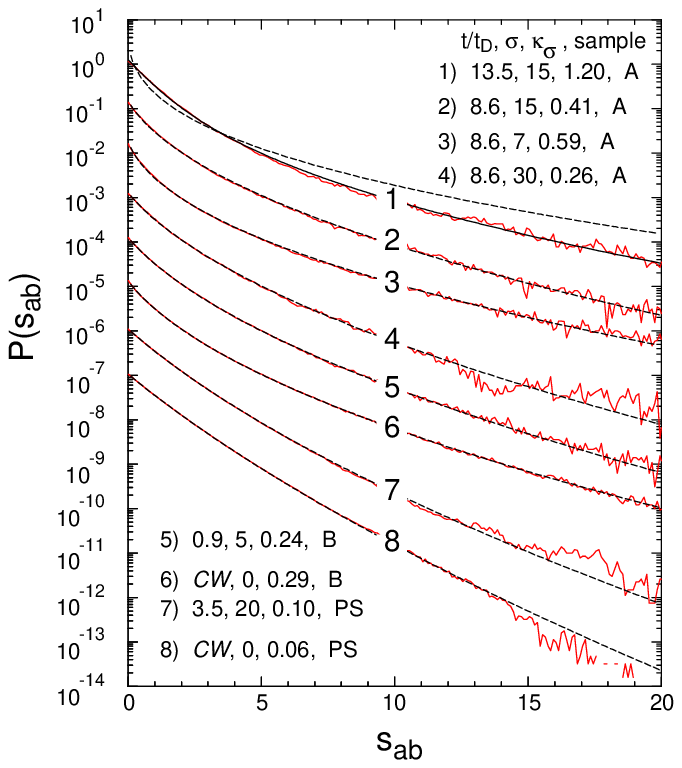}
\caption{Dynamic and steady-state (\textit{CW}) probability
distributions of normalized transmitted intensity (solid curves)
with the values of $t/t_D$, $\sigma$, and $\kappa_{\sigma}$, and
sample type indicated. The dashed curves are from Eqs.~(1-3). In
the case of distribution (1) of long delay time, noise accounts
for nearly half the measured field. The curve through the data is
obtained by transforming accordingly Eqs.~(1-3) to include the
effect of noise. In the cases other than (1), noise is
disregarded.}
\end{figure}

The correlation function of the normalized transmitted field with
shift in polarization angle $\Delta\theta$, $F_E(\Delta\theta)$,
is shown in Fig.~2a for the two delay times indicated by the
vertical lines in the inset for the incident pulse of bandwidth
$\sigma$=5 MHz and for monochromatic excitation. All functions are
well described by the steady-state result,
$F_E(\Delta\theta)$=$\cos(\Delta\theta)$ \cite{Polar04,Freund90}.
The corresponding intensity correlation functions are shown in
Fig.~2b. These have the form,
$C_{\sigma}(\Delta\theta,t)$=$F(\Delta\theta)$+$\kappa_{\sigma}(t)[F(\Delta\theta)$+$1]$,
with $\kappa_{\sigma}(t)$=$C_{\sigma}(90,t)$.

In order to study the dynamics of spatial correlation,
measurements are taken at 50 points separated by 1.06 mm along a
line centered on the tube axis on the output surface of a random
sample of Polystyrene spheres \cite{Patrick02}. A 3-mm antenna is
aligned perpendicular to the line of displacement. The Polystyrene
spheres of refractive index 1.59 are packed at a volume filling
fraction of 0.52 in a copper tube with $L$=100 cm. Measurements
are made in 1380 sample realizations over the frequency range
17.2-17.8 GHz in steps of 0.625 MHz. In this sample,
$\kappa$=0.06.

The correlation function of the normalized transmitted field with
displacement $\Delta r$, $F_E(\Delta r)$, is shown in Fig.~3a for
the two delay times indicated in the inset following pulsed
excitation with $\sigma$=20 MHz and for monochromatic excitation.
The overlap of these curves indicates that the dynamic $F_E(\Delta
r)$ is independent of $t$ and identical with the steady-state
field correlation function. The corresponding intensity
correlation functions, shown in Fig.~3b, have the form,
$C_{\sigma}(\Delta r,t)$=$F(\Delta
r)$+$\kappa_{\sigma}(t)[F(\Delta r)$+$1]$, with
$\kappa_{\sigma}(t)$ determined from the residual correlation at
displacements $\Delta r>3.5$ cm, except for $t/t_d$=4.8 when
$\kappa_{\sigma}(t)$ is determined from the relation,
var$(s_{ab}(t))_{\sigma}$=$C_{\sigma}(0,t)$=1+$2\kappa_{\sigma}(t)$.

The time-resolved probability distributions of normalized
transmitted intensity, $P(s_{ab}(t))$, are shown in Fig.~4 for the
three samples considered, for various values of delay time and
pulse bandwidth. The steady-state intensity distributions for
Sample B and for the Polystyrene sample are also shown. Apart from
the uppermost curve, each of the curves is displaced by a multiple
of one decade for clarity of presentation. The dashed curves are
obtained from the expressions for the steady-state intensity
distribution $P(s_{ab})$ \cite{Rossum,Kogan} in the limit,
$g$$\gg$1, and in the absence of absorption, but with
$2/3\kappa_{\sigma}$ substituted for $g$,
\begin{equation}
P(s_{ab})=\int_{0}^{\infty}\frac{ds_a}{s_a}P(s_a)\exp(-s_{ab}/s_{a})\, ,
\end{equation}
with
\begin{equation}
P(s_{a})=\int_{-i\infty}^{i\infty}\frac{d\upsilon}{2\pi i}\exp{\![\upsilon s_{a}-\Phi(\upsilon)]},
\end{equation}
where
\begin{equation}
\Phi(\upsilon)=(2/3\kappa_{\sigma})
\ln^{2}\!\left(\sqrt{1+3\upsilon\kappa_{\sigma}/2}+\sqrt{3\upsilon\kappa_{\sigma}/2}\right).
\end{equation}
The values of $\kappa_{\sigma}$ for Sample B and for the
Polystyrene sample are obtained from the measured intensity
correlation functions in Figs.~2 and 3. The values of
$\kappa_{\sigma}$ for Sample A are determined from the relation,
$C_{\sigma}(0,t)$=1+$2\kappa_{\sigma}(t)$.
Excellent agreement with measured results is obtained in all
cases. These encompass both steady state and dynamic propagation,
in the presence of strong and weak absorption, in the weak as well
as the strong correlation regimes, in which $\kappa_{\sigma}$
exceeds its value at the Anderson localization threshold of 2/3.

In conclusion, the time-resolved intensity correlation function
has a universal form in terms of $\kappa_{\sigma}(t)$ and the
square of the time-independent field correlation function, $F$. At
the same time, the probability distribution of intensity is
determined exclusively by $\kappa_{\sigma}(t)$. In the limit,
$\sigma$$\to$0, $\kappa_{\sigma}$$\to$$\kappa$, yielding
steady-state statistics. Thus, the time-varying degree of
correlation, $\kappa_{\sigma}(t)$, is the controlling function of
mesoscopic statistics. It is, therefore, of prime importance to
explore the possibility of a universal formulation of the time
variation of the degree of correlation and its relationship to
spatial localization.

We thank P.~Sebbah for helpful suggestions. This research is
sponsored by the National Science Foundation (DMR0205186) and U.S.
Army Research Office (DAAD190010362).

\end{document}